\documentclass[twocolumn,prl]{revtex4}
\usepackage{graphicx}
\begin{document}


\title{Critically damped quantum search}
\author{Ari Mizel}
\affiliation{Science Applications International Corporation, 4001 N. Fairfax Drive Arlington, VA 22203}
\email{ari.m.mizel@saic.com}
\date{\today}
\begin{abstract}
Although measurement and unitary processes can accomplish any quantum evolution in principle, thinking in terms of dissipation and damping can be powerful.  We propose a modification of Grover's algorithm in which the idea of damping plays a natural role.  Remarkably, we have found that there is a critical damping value that divides between the quantum $O(\sqrt{N})$ and classical $O(N)$ search regimes.  In addition, by allowing the damping to vary in a fashion we describe, one obtains a fixed-point quantum search algorithm in which ignorance of the number of targets increases the number of oracle queries only by a factor of 1.5.
\end{abstract}

\maketitle

Dissipation has generally been regarded as a destructive foe in the arena of quantum mechanics, ruining quantum effects and greatly complicating theory.  This is particularly true in the case of quantum information science, which depends on maintaining delicate quantum superpositions and entanglement.  On the other hand, dissipation has many productive uses in the classical regime, from automobile shocks to toaster ovens.  One therefore wonders whether dissipation might be employed constructively in the quantum information context \cite{constructive}.

We propose here a natural application of dissipation in the quantum search algorithm.  This algorithm, due to Grover \cite{Grover97}, is a mainstay of quantum information science.  Given an unsorted database of $N$ items of which $M$ are target items, the quantum search locates one of the target items with $O(\sqrt{N/M})$ queries of an oracle.  A classical search would require $O(N/M)$ queries.  A well known vulnerability of the quantum search is the need for prior knowledge of the value of $M$ \cite{Boyer98,Grover98}.  In the absence of such knowledge, the quantum search is not robust, producing results that oscillate between target and non-target items.  In this Letter, we show that introducing dissipation into the search algorithm can damp out these oscillations.  Strikingly, we find that a critical damping value emerges from the theory that divides between the quantum regime (low dissipation, , $O(\sqrt{N/M})$ queries) and the classical regime (high dissipation, , $O(N/M)$ queries).  

Although recently other fixed-point quantum searches have been developed \cite{Grover05,Tulsi06,Grover06}, they are not designed to preserve the signature $O(\sqrt{N/M})$ behavior of the quantum search.  When the damping is chosen appropriately, the dissipative approach maintains the $O(\sqrt{N/M})$ quantum behavior.  Furthermore, when the damping is allowed to vary in a manner we describe, ignorance of $M$ costs only a factor of $1.5$ in the number of oracle calls, which is a very low overhead compared to other ways of handling ignorance of $M$.  Overall, our results convincingly demonstrate the productive use of dissipation in quantum algorithms and provide an example where the appropriate amount of dissipation emerges explicitly from the theory.

We briefly review Grover's quantum search \cite{Nielsen00}.  Given an unsorted database of $N$ items, one is charged with the task of finding any one of $M$ target items dispersed throughout the database.  The tool to probe the database is an oracle that indicates whether a given item is a target or not.  The algorithm begins by placing the state $\left| \psi \right>$ of the system into an equal superposition of all database states $\sum_{s = 1,...,N} |s>/\sqrt{N} = \cos \theta/2 \left| \alpha \right> + \sin \theta/2 \left| \beta \right>$.  Here, we have defined $\left| \alpha\right>$ to be an equal superposition of all $M-N$ non-target states, $\left| \beta \right>$ to be an equal superposition of all $M$ target states, and $\sin \theta/2 = \sqrt{M/N}$.  It is convenient to define Pauli operators in the 2-dimensional Hilbert space of $\left| \alpha\right>$ and $\left| \beta \right>$ by $X = \left|\beta\right>\left< \alpha\right|  +\left|\alpha\right>\left< \beta\right|$, $Y = i \left|\beta\right>\left< \alpha\right|  - i\left|\alpha\right>\left< \beta\right|$, and $Z = \left|\alpha\right>\left< \alpha\right| - \left|\beta\right>\left< \beta\right|$.  In terms of these operators, the oracle is just $Z$.  Grover developed a series of gates, consisting of one call to the oracle $Z$ followed by one "inversion about the mean" $E$, that produces the rotation $G = E Z = \exp(-i \theta Y)$.  After $R$ applications of $G$, the system has rotated to $\cos(2R+1)\theta/2 \, \left| \alpha \right> + \sin(2R+1)\theta/2 \, \left| \beta \right>$.  If $\theta$ is not too big ($M \ll N$), and one chooses $R$ to be near $R(M) \equiv \lceil \arccos(\sqrt{M/N})/\theta \rceil$, then the final state will be close to the target $\left| \beta \right>$.  Clearly, if one is ignorant of $M$ and keeps applying $G$ past $R \sim R(M)$, the system will rotate past $\left| \beta \right>$.

In terms of the system density matrix, the evolution $\rho ^\prime = G \rho G^\dagger$ implies
\begin{equation}
\label{undampedrho}
\left[\begin{array}{c} \mathbf{Tr}(\rho^\prime X) \\ \mathbf{Tr}(\rho^\prime Z)  \\ \mathbf{Tr}(\rho^\prime) \end{array}\right] = \left[\begin{array}{rrr} \cos 2\theta &\sin 2\theta & 0 \\ -\sin 2\theta &\cos 2\theta & 0 \\ 0& 0 & 1 \end{array}\right]\left[\begin{array}{c} \mathbf{Tr}(\rho X) \\ \mathbf{Tr}(\rho Z)  \\ \mathbf{Tr}(\rho) \end{array}\right]
\end{equation}
where the initial density matrix $\rho = \left| \psi \right> \left<\psi \right|$ satisfies $\left[\mathbf{Tr}(\rho X) \,\, \mathbf{Tr}(\rho Z) \,\, \mathbf{Tr}(\rho) \right]^T = \left[\sin \theta \,\, \cos\theta \,\, 1\right]^T$.

So far, we have simply reviewed Grover's algorithm, pointing out how it assumes prior knowledge of the number of targets $M$.  Suppose that we do not know $M$.  How can we find a target reliably?  We propose introducing damping to slow the rotation in the $\left| \alpha \right>$, $\left| \beta \right>$ plane as the system approaches the target state.

We append an external spin to the system, taking $\left| \psi\right>$ to $\left| \psi\right> \left| \downarrow \right>$.  Let the external spin Pauli operators be $S_x, S_y, S_z$.  This external spin will serve to indicate the proximity of the system to the target state.  We replace the Grover rotation $G$ with
\begin{equation}
\label{U}
U = \left[G \frac{1-S_z}{2} + \frac{1+S_z}{2}\right] \left[e^{-i \phi S_y} \frac{1-Z}{2} + \frac{1+Z}{2}\right].
\end{equation}
First, $\left[e^{-i \phi S_y} (1-Z)/2 + (1+Z)/2\right]$ calls the oracle and flips the external spin if $\left| \psi\right>$ has reached the target state ($Z=-1$).  Second, $\left[G (1-S_z)/2 + (1+S_z)/2\right]$ applies a Grover rotation only if the external spin has not flipped ($S_z = -1$).  Thus, the external spin stifles the Grover rotation as the target is approached, effectively damping the system.  Although it appears at first that $U$ requires 2 controlled oracle calls, in fact it can be written using one controlled oracle call as
\[
U = \left[E \frac{1-S_z}{2} + \frac{1+S_z}{2}\right] e^{i\phi S_y/2} \left[Z \frac{1-S_z}{2} + \frac{1+S_z}{2}\right]e^{-i\phi S_y/2}.
\]

Assume we follow application of $U$ with measurement of the external spin, and repeat these two steps until the external spin has flipped.  The density matrix of the system will have the form $\rho \left| \downarrow \right> \left< \downarrow \right|$ until the external spin flips.  The iteration produces the following effect on $\rho$:
\begin{widetext}
\begin{equation}
\label{dampedrho}
\left[\begin{array}{c} \mathbf{Tr}(\rho^\prime X) \\ \mathbf{Tr}(\rho^\prime Z)  \\ \mathbf{Tr}(\rho^\prime) \end{array}\right] = \left[\begin{array}{rrr} \cos 2\theta \cos \phi &\sin 2\theta \frac{1+\cos^2 \phi}{2}& \sin 2\theta \frac{1-\cos^2 \phi}{2} \\ -\sin 2\theta\cos \phi &\cos 2\theta\frac{1+\cos^2 \phi}{2} & \cos 2\theta\frac{1-\cos^2 \phi}{2} \\ 0& \frac{1-\cos^2 \phi}{2} & \frac{1+\cos^2 \phi}{2} \end{array}\right]\left[\begin{array}{c} \mathbf{Tr}(\rho X) \\ \mathbf{Tr}(\rho Z)  \\ \mathbf{Tr}(\rho) \end{array}\right]
\end{equation}
\end{widetext}
Note that, since there is some probability at each iteration that the external spin will flip, $\mathbf{Tr}(\rho)$ can decrease.  

The value of $\phi$ determines the amount of damping of the iteration.  When $\phi = 0$, there is no damping, and we recover Grover's quantum search.  When $\phi = \pi/2$, the damping is strongest -- the external spin essentially acts as a pointer for a full measurement of $Z$, the system undergoes collapse at each oracle call, and the search is nearly classical.  This is the limit discussed in \cite{Tulsi06,Grover06}.  To choose the optimal value of $\phi$ for our purposes, we consider the eigenvalues of the matrix in (\ref{dampedrho}).  As Fig. 1 shows, for small $\phi$, when $U$ is nearly $G = \exp(-i \theta Y)$ and (\ref{dampedrho}) is nearly (\ref{undampedrho}), there are two complex conjugate eigenvalues $\sim \exp(\pm i 2\theta)$ associated with the rotation and one eigenvalue $\sim 1$ that keeps $\mathbf{Tr} (\rho)$ nearly constant.  When $\phi$ is large, there are three real eigenvalues.  In the extreme case $\phi = \pi/2$ (not shown in Fig. 1), two eigenvalues are 0 and one equals $\cos^2\theta$.  The eigenvalue associated with $\mathbf{Tr}(\rho X)$ is 0 since the measurement destroys the coherence.  The eigenvalue associated with $\mathbf{Tr}(\rho (1-Z)/2)$ is 0 since $\rho$ is associated with the non-target part of the system for which $Z=1$.  The remaining eigenvalue is $\cos^2 \theta$ because a projection to the non-target state $\mathbf{Tr}(\rho (1+Z)/2)$ followed by a Grover rotation through $\theta$ returns a non-target state with probability $\cos^2\theta$.

\begin{figure}
\includegraphics[width=\linewidth]{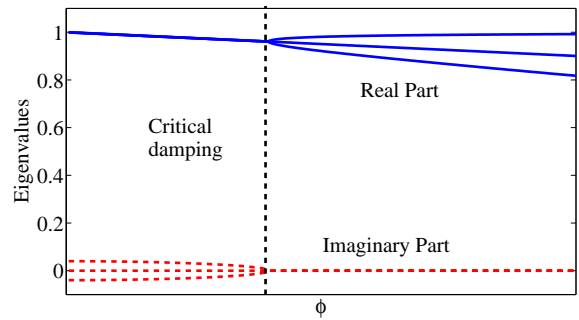}
\caption{Real and imaginary parts of the 3 eigenvalues of the matrix (\ref{dampedrho}) as a function of the damping $\phi$ near the critical damping.  Since the qualitative appearance of graph is independent of $M$ for $M \ll N$, we have avoided putting numerical values on the $\phi$ axis specific to a particular $M$.  The critical damping occurs at $\cos \phi = (1-\sin \theta)/(1+\sin\theta) = (N-2\sqrt{M(N-M)})/(N+2\sqrt{M(N-M)})$.}
\end{figure}

Remarkably, there is a critical damping defined by $\cos \phi = (1-\sin \theta)/(1+\sin\theta)$ at which all three eigenvalues are $\cos \phi$.  At this critical damping, all three components $\mathbf{Tr}(\rho X)$, $\mathbf{Tr}(\rho Z)$, and $\mathbf{Tr}(\rho)$ appearing in the map (\ref{dampedrho}) tend to be suppressed since all the eigenvalues have magnitude under 1.  Thus, $1 - \mathbf{Tr}(\rho)$, the probability that the target state has been found and the external spin has flipped, tends to increase.  Fig. 2 exhibits the special character of the critical damping value.  It depicts the average number of oracle calls to find the target as a function of $N$ and $\phi$ in the case $M=1$.  As the damping increases, the Grover rotation shrinks, so the number of oracle queries tends to increase.  For small damping, the average number of oracle queries to find the target has the quantum behavior $O(\sqrt{N})$, while for large damping, the number of queries goes like the classical expectation $O(N)$.  However, for the critical value of the damping, there is a distinct valley that separates these two regimes.  (The number of oracle calls on the z-axis of Fig. 2 is computed assuming the following reasonable strategy.  Knowing $N$ and $\phi$, one plans to execute $R$ calls to $U$.  If the external spin flips before the $R$ calls are complete, a target has been found with certainty and no more calls are needed.  Otherwise, the $R$ calls to $U$ are followed with one query to the oracle to determine if a target has been found.  If not, one starts the procedure over.  By fixing a judicious choice of $R$, one minimizes the expected number of oracle queries.  This minimum is plotted on the z-axis in Fig. 2.)

\begin{figure}
\includegraphics[width=\linewidth]{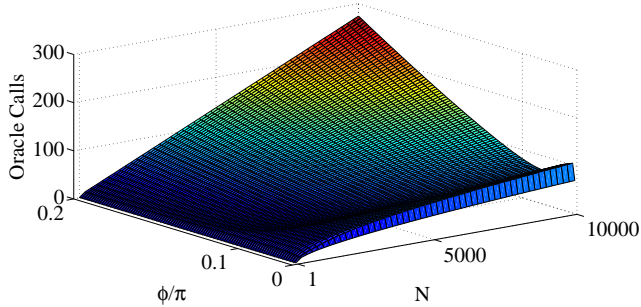}
\caption{Average number of oracle calls to find the target for $M=1$ as a function of $N$ and $\phi$.  Note the linear dependence on $N$ for large $\phi$, the $\sqrt{N}$ dependence for small $\phi$, and the valley separating these two regions at the critical damping.}
\end{figure}

Having pointed out the existance of a critical damping value and noted its special character, we now consider how the damping can be used in the case of unknown $M$.  The qualitative appearance of Fig. 1 is maintained for general $M \ll N$, but the value of the critical damping depends upon $M$ through $\theta$.  In the absence of knowledge of $M$, suppose we assume the worst-case scenario of fewest targets, $M=1$, in which the damping $\cos \phi = (1-\sin \theta)/(1+\sin \theta) = (N-2\sqrt{M(N-M)})/(N+2\sqrt{M(N-M)}) = (\sqrt{N-1}-1)^2/(\sqrt{N-1}+1)^2$ is weakest.  Fig. 3 shows the progress of the algorithm given this choice of $\phi$, for $N=10,000$ with $M=1$ and with $M=40$.  The effect of the dissipation is evident; the oscillations into and out of the target state are effectively damped even for $M$ substantially greater than 1.  The value of $Tr(\rho Z)$ tends to zero, and with high probability the external spin flips.  Since the external spin signals success, it is not necessary to know $M$ to run the search effectively.

\begin{figure}
\includegraphics[width=\linewidth]{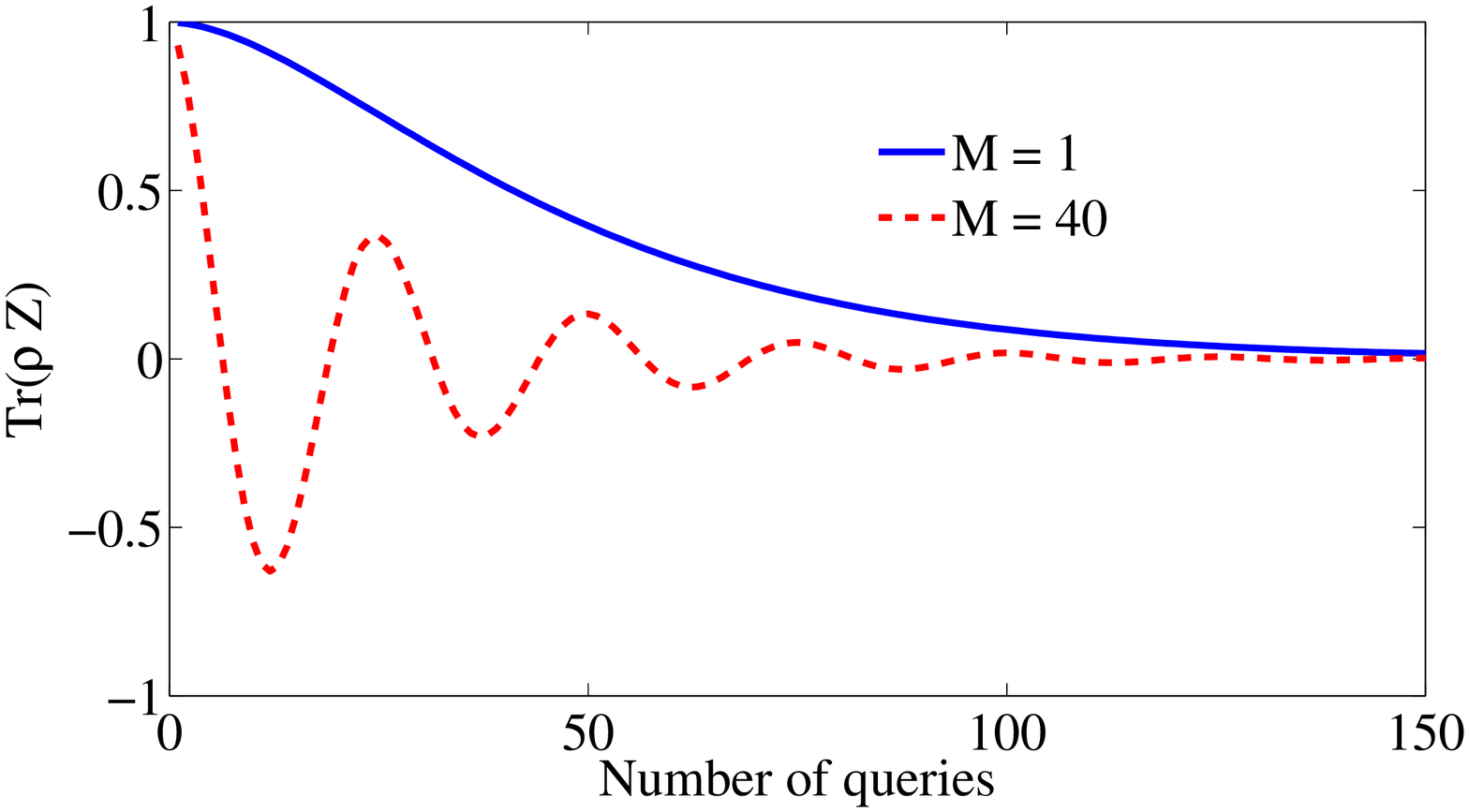}
\caption{Damping of oscillations of $Tr(\rho Z)$.  They approach 0 rather than -1 since $\rho$ is only the part of the density matrix for which the external since has not flipped.}
\end{figure}

Although the choice $\cos \phi = (\sqrt{N-1}-1)^2/(\sqrt{N-1}+1)^2$ is effective for a broad range of $M$ values near $M=1$, it is not effective when the unknown value of $M$ happens to be very large, on the order of $N/2$.  For such large values of $M$, the angle $\theta$ is much greater than this choice of $\phi$, and the system approaches the target state long before the external spin flips to indicate success.  To adapt our algorithm so that the spin can flip early when appropriate, we allow $\phi$ to change from iteration to iteration.  In the first application of $U$, there is large damping $\phi = \pi/2$ in case $M$ is large.  If the external spin has not flipped after this iteration of $U$, this is taken as evidence that $M$ must be somewhat smaller, so the damping is decreased.  For concreteness, for iteration $n>1$, we set $\cos \phi_n = (1-\sin(\pi/2n))/(1+\sin(\pi/2n))$, which should be somewhat near the critical damping for the $M$ satisfying $R(M) \sim n$.  Although this choice of $\phi_n$ has not been optimized, we find that it yields good behavior.

The results are shown in Fig. 4, which compares the average number of iterations before the external spin flips to the average number of iterations of the undamped Grover's algorithm assuming prior knowledge of $M$. (The average number of iterations of the undamped Grover's algorithm is determined as follows. First, one computes $p(R)$, the probability of finding a target after $R$ calls to the oracle given $N$ and $M$.  One then minimizes over $R$ the average number of iterations $(R+1)/p(R) = (R+1) p(R) + 2(R+1) p(R) (1-p(R)) + 3(R+1) p(R) (1-p(R))^2 + ...$ assuming that one performs a verification call to the oracle after the $R$ iterations and then repeats the whole procedure if the verification turns out negative.)  In the worst case shown, ignorance of $M$ leads to an extra factor of roughly $1.5$ oracle calls, although occasionally damping can actually decrease the number of oracle calls.  This damping method of coping with ignorance of $M$ compares very favorably with other methods such as successively applying $2^n$ Grover iterations and measuring, which behaves quite erractically as a function of $M$ sometimes costing an extra factor of 30 or more oracle calls, or quantum counting \cite{Boyer98}, which yields the value of $M$ but at a factor far in excess of $1.5$ if $M$ is determined with reasonable accuracy.

\begin{figure}
\includegraphics[width=\linewidth]{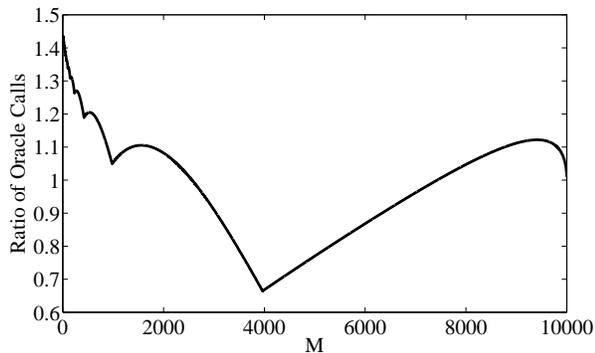}
\caption{Average number of calls to oracle for damped search with varying damping $\cos \phi_n = (1-\sin(\pi/2n))/(1+\sin(\pi/2n))$ as a function of $M$.  We take $N=10,000$ and assume ignorance of $M$.  For each $M$, the damped search result is divided by the average number of calls to the oracle for the undamped quantum search assuming that $M$ is known.  Note that ignorance of $M$ costs at most a factor of $1.5$ in oracle calls.}
\end{figure}

So far, the damping has been effected "artificially" using a single ancilla spin that is subjected to unitary evolution with the system and then measured.  This could be the best way to proceed in an actual quantum computer, but it is also possible to produce the damping in a more traditional fashion by coupling the system to a bath.  To show this, we first append a flag qubit to the system to signal when the target has been reached, taking $\left| \psi\right>$ to $\left| \psi\right> \left| 0 \right>$.  The flag qubit operators are denoted $\sigma_x$, $\sigma_y$, and $\sigma_z$.  Next, we introduce a low-temperature bath, modeled as a collection of spins with operators $\{S_{x,i},S_{y,i},S_{z,i}\}$.  We can achieve the same results as (\ref{U}) by revising it to
\begin{eqnarray}
\label{bathU}
U&=&[G \frac{1-\sigma_z}{2}+\frac{1+\sigma_z}{2}] \\
& &\prod_i [\sigma_x \frac{1+S_{z,i}}{2}+\frac{1-S_{z,i}}{2}] \nonumber \\
& &[(e^{-i\phi S_{y,i}}-1)\frac{1-Z}{2}\frac{1-\sigma_z}{2}+1] \nonumber
\end{eqnarray}
The similarity to (\ref{U}) is evident, but there is an extra operator $[\sigma_x (1+S_{z,i})/2+(1-S_{z,i})/2]$ that flips the flag qubit if the bath spin has flipped.  We can write (\ref{bathU}) as $U = \exp(-i\theta Y (1-\sigma_z)/2) \prod_i [\sigma_x \frac{1+S_{z,i}}{2}+\frac{1-S_{z,i}}{2}] \exp(-i\phi S_{y,i}(1-Z)(1-\sigma_z)/4) = \exp(-i\theta Y (1-\sigma_z)/2) \prod_i \exp(\phi (1-Z)/2 (\sigma_+ S_{+,i} - \sigma_- S_{-,i}))[\sigma_x \frac{1+S_{z,i}}{2}+\frac{1-S_{z,i}}{2}]$.  Assuming that the bath is cold so that $S_{z,i} = -1$ initially, the last factor can be removed.  Now, one can think of Grover's algorithm as evolution under a Hamiltonian \cite{Farhi98}, and since the Grover rotation has the form $G = \exp(-i \theta Y)$ it follows that $H=Y$.  To add damping to the Hamiltonian, we are motivated by the form of $U$ to write $H = Y (1-\sigma_z)/2 + i(\phi/\theta) (1-Z)/2 (\sigma_+ \sum_i S_{+,i} + \sigma_- \sum_i S_{-,i}) + H_{bath}$.  Assuming that $H_{bath}$ allows us to make a Markovian approximation, we write a Lindbladt equation \cite{Breuer02} for the portion $\rho$ of system density matrix for which the flag qubit has not flipped: $\dot{\rho} = -i [Y,\rho] - C (1-Z)\rho/2  - C \rho (1-Z)/2$.  Here, $C$ depends on $\phi/\theta$ and on spin correlation functions of the bath.  This Lindbladt equation leads to dynamics analogous to those resulting from (\ref{U}).   

In summary, we have introduced damping into Grover's search to mitigate over-rotation past the target states when the number of target states is unknown.  A critical damping value has emerged that divides between the classical and quantum regimes.  Tuning the damping appropriately permits quantum search without knowledge of $M$ with only a factor of $1.5$
overhead.  We have presented one promising application of dissipation in quantum information, but others can be identified, suggesting that this is an exciting avenue for further study.

The author acknowledges helpful discussions with Andrew Cross, Daniel Gottesman, Patrick Hayden, and Kevin Obenland.

\end{document}